\documentclass[aps,nofootinbib,showpacs,preprintnumbers,amsmath,amssymb]{revtex4}
\usepackage{epsfig}
\usepackage{bm}

\begin{document}
\begin{flushright} hep-ph/0608203 v2, preprint USM-TH-194 \end{flushright}

\title{Lepton flavor violation in muonium decay and muon
colliders\\ in models with heavy neutrinos}

\author{G.~Cveti\v c}
  \email{gorazd.cvetic@usm.cl}
\author{C.~Dib}
  \email{claudio.dib@usm.cl}
\affiliation{Dept.~of Physics, Universidad T\'ecnica
Federico Santa Mar\'{\i}a, Valpara\'{\i}so, Chile}
\author{C.~S.~Kim}
   \email{cskim@yonsei.ac.kr}
\affiliation{Department of Physics, Yonsei University,
Seoul 120-749, Korea}
\author{J.~D.Kim}
    \email{jade@phya.snu.ac.kr}
\affiliation{Department of Physics, Seoul National University,
Seoul 151-742, Korea}


\begin{abstract}
\noindent We study the lepton-flavor-violating reaction $\mu^+ e^-
\to e^+ e^-$ within two extensions of the standard model that
include heavy neutrinos. The reaction is studied in the low energy
limit in the form of muonium decay $M\to e^+ e^-$ and in the high
energy regime of a muon collider. The two theoretical models we
consider are: model I, a typical see-saw model that violates
lepton flavor and number by inclusion of extra right handed
neutrinos, and model II, a variant where lepton number is
conserved and which includes extra right handed as well as left
handed neutrinos, singlets under the gauge group. We find for
muonium decay into $e^+e^-$ the extremely small result $Br(M\to
e^+ e^-) < 10^{-19}$ in both scenarios. Alternatively, for $\mu^+
e^-$ collisions up to $\sqrt{s}\sim 50$ GeV we find $\sigma(\mu^+
e^- \to e^+ e^-)< 10^{-5}$ fb, while for energies above the $W^+
W^-$ threshold we find $\sigma(\mu^+ e^- \to W^+ W^-)$ up to $\sim 1$ fb.

\end{abstract}

\maketitle

\section{Introduction}

A current issue in particle physics is the understanding of the
neutrino sector. It is by now consistent with experimental results
to assert that at least some of the standard neutrinos must have
small (but non-zero) masses and these mass states should mix under
the weak interactions\cite{Fogli}. As a consequence, the lightness
of the standard neutrinos ($m_\nu \lesssim 1$ eV) compared to the
charged leptons ($\gtrsim$ MeV) remains to be understood. If
neutrinos are of a Dirac nature, non-zero masses can appear in the
standard model (SM) by inclusion of (sterile) right-handed
neutrinos\cite{Schechter}. If neutrinos are of a Majorana nature
--a more general case--, more appealing solutions exist in the
context of extended gauge structures. An interesting solution is
provided by the so-called seesaw mechanism within $SO(10)$ or
left-right symmetric models of interactions. In conventional
seesaw models, the effective light neutrino masses are within the
scales of eV to MeV via a relation involving the hierarchy between
very large Majorana masses and Dirac masses comparable to those of
the charged leptons\cite{Seesaw}. Another possible solution was
investigated in the framework of heterotic superstring
models~\cite{E6} with $E_6$ symmetry or certain scenarios of
$SO(10)$ models~\cite{SO10a}, where the low--energy effective
theories include new left--handed and right--handed neutral
isosinglets and assume conservation of total lepton number in the
Yukawa sector.

One avenue to study the neutrino sector is to consider lepton
flavor violating processes (LFV) of standard particles, which are
absent in the SM and depend on neutrino masses and mixings. Free
lepton decays like $\mu\to e\gamma$ $\mu\to 3e$, $\tau\to 3\mu$,
etc.\ are among the most direct LFV processes and they have been
studied in detail \cite{Lee}-\cite{Kageyama},\cite{CDKK}. Other
ways like muon-to-electron conversion in nuclei and
muonium-to-antimuonium conversion are also possibilities
\cite{muonium,Willmann,CDKK2}. Yet another possibility is to
consider lepton-lepton scattering or bound state decays (e.g.
muonium decay). These modes contain the same underlying physics as
free lepton decays, but differ in the experimental conditions and,
from the phenomenological viewpoint they are sensitive to other
combination of parameters because of their different kinematical
regime.

In this work we study the process $\mu^+ e^- \to e^+ e^-$ at low
energy in the form of muonium decay (muonium is a non-relativistic
$\mu^+ e^-$ bound state) and at high energy in the form of $\mu^+
e^-$ collisions. Muonium is formed when a $\mu^+$ slows down
inside a material and captures an electron, forming a bound state
due to Coulomb attraction. The fraction of muons that actually end
up as muonium depends very much on the material, ranging from
small fractions in some to nearly 100\% in others. Most muonia
decay as $M\to e^+e^-\bar\nu_\mu \nu_e$, which is simply a muon
decay with an electron as spectator \cite{Czarnecki}. A more rare
mode, albeit standard, is annihilation $M\to\bar\nu_\mu \nu_e$,
whose rate is $\sim 10^{-11}$ times smaller. Here we study the
mode $M\to e^+ e^-$, which violates lepton flavor and is thus
forbidden in the SM. From our point of view, muonium decay is
interesting because it is a non-relativistic $\mu^+ e^-$ process,
and thus depends on different combinations of parameters than in
high energy processes. Complementary to it, we also study the
collision $\mu^+ e^-\to e^+e^-$ at high energy. This process
clearly requires a muon collider. We will not be exhaustive in the
study of new physics at muon colliders but focus on this
particular channel within the LFV models previously mentioned. In
section II we give a brief review of the seesaw type models in
question; in section III we describe the generic amplitudes for
$\mu^+ e^-\to e^+ e^-$ processes within these models; section IV
contains the application to muonium decay, section V the
application to high energy collisions and section VI summarizes
the results and conclusions.

\section{Two Models for heavy neutrinos}
\label{sec:models}

Here we will briefly summarize the two models under consideration.
Further details were given in previous works\cite{CDKK, CDKK2}.
The main features of these models are (i) they contain heavy
neutrinos, (ii) in the mass eigenstate basis, the heavy neutrinos
couple with charged leptons of mixed flavor under weak
interactions (just like Cabbibo-Kobayashi-Maskawa mixing in the
quark sector), and (iii) the neutrinos are generally of Majorana
type.

\noindent{\bf Model I:}\ \ This is a usual seesaw model
\cite{Seesaw}: it is the standard model with $N_L$ generations
({\it i.e.} left-handed neutrinos $\nu_{i L}$) and an equal number
$N_H=N_L$ of right-handed neutrinos ${\widetilde \nu}_{i R}$,
singlets under the gauge group.  The neutrino mass terms, after
gauge symmetry breaking, are:
\begin{equation}
-{\cal L}^{\nu}_{\rm mass}=
\frac{1}{2} \left( {\overline {{\nu}_L}},
{\overline {{\widetilde \nu}_R^{ \  c}}}   \right)
{\cal M}
\left(
\begin{array}{c}
\nu_L^{ \  c} \\ {\widetilde \nu}_R
\end{array} \right) +  {\rm h.c.}
\label{Lmass}
\end{equation}
The superscript $c$ at some of the fields denotes
charge-conjugated fields $\psi^c = {\cal C} {\overline \psi}^T$,
where ${\cal C} = - i \gamma^2 \gamma^0$ in the Dirac
representation. The $2 N_L \times 2 N_L$--dimensional matrix
${\cal M}$ has a seesaw block form:
\begin{equation}
{\cal M} = \left(
\begin{array}{cc}
0 & m_D \\ m_D^T & m_M
\end{array}
\right), \quad m_D= \left(
\begin{array}{cc}
a & b \; e^{i \delta_1}  \\ c \; e^{i \delta_2}&  d
\end{array}
\right), \quad m_M= \left(
\begin{array}{cc}
M_1 & 0 \\ 0 & M_2
\end{array}
\right)  \ , \label{Dm1}
\end{equation}
The (real) parameters $a$, $b$, $c$, $d$ in the Dirac mass matrix
$m_D$ are smaller than the Majorana mass parameters $M_1$ and $M_2$
($M_2 \geq M_1 \agt 100$ GeV) by at least an order
of magnitude.

The squared ratio between the Dirac mass ($m_D$) and the Majorana
mass ($m_M$) scales gives the order of magnitude of the
heavy-to-light neutrino mixings $(s_L^{\nu_l})^2 \equiv \sum_{h} |
U_{h l} |^2 \sim |m_D|^2/|m_M|^2$. The physical light neutrino
masses are of the order of $m_{\nu_{light}}\!\sim\!m^2_D/m_M$. The
very low experimental bounds $m_{\nu_{light}} \stackrel{<}{\sim}
1$ eV impose in Model I severe constraints on the $|m_D| \ll
|m_M|$ hierarchy required. The present experimental bounds on the
heavy-to-light mixing parameters $(s_L^{\nu_l})^2 \sim
|m_D|^2/|m_M|^2$ ($\stackrel{<}{\sim} 10^{-2}$, see below) present
another set of constraints on the model.

\noindent{\bf Model II:}\ \  This model contains an equal number
$N_L$ of left-handed ($S_{i L}$) and right-handed (${\widetilde
\nu}_{i R}$) neutral singlets~\cite{E6,SO10a}. Furthermore, the
form of the mass matrix ${\cal M}$ leads to the conservation of
total Lepton Number, although lepton flavor mixing is still
possible. The neutrino mass terms, after electroweak symmetry
breaking, have the form
\begin{equation}
-{\cal L}^{\nu}_{\rm mass}=\frac{1}{2}
\left( {\overline {{\nu}_L}},{\overline {{\widetilde \nu}_R^{ \ c}}},
{\overline {{S}_L}} \right)
{\cal M}
\left(
\begin{array}{c}
{\nu_L^{ \ c}} \\ {\widetilde \nu}_R \\ S_L^{ \ c}
\end{array}
\right)
+ {\rm h.c.} \ , \qquad
{\cal M}= \left(
\begin{array}{ccc}
0 & m_D & 0\\ m_D^T & 0 & m_M^T \\ 0 & m_M & 0
\end{array} \right) \ ,
\label{M-model2}
\end{equation}
and $m_D$ and $m_M$ are given in Eq.~(\ref{Dm1}). The mass matrix
${\cal M}$ is $(N_L +N_H)\times (N_L+N_H)$-dimensional, $N_H =
2N_L$ (the Dirac block $m_D$ is $N_L \times N_L$--dimensional).
The model, in its unperturbed form, predicts for each of the $N_L$
generations a massless Weyl neutrino and two degenerate Majorana
neutrinos ~\cite{BRV,Gonzalez-Garcia:1989rw} ($N_H = 2 N_L$).
Therefore, the seesaw-type restriction
$m_{\nu_{light}}\!\sim\!m^2_D/m_M$ of Model I is not present in
Model II in its unperturbed form. However, present experimental
bounds on heavy-to-light mixing parameters $(s_L^{\nu_l})^2 \sim
|m_D|^2/|m_M|^2$ ($\stackrel{<}{\sim} 10^{-2}$) do impose a
certain level of hierarchy $|m_D| < |m_M|$ between the Dirac and
Majorana mass sector, but is in general significantly weaker than
in Model I. Nonzero masses of the $N_L$ light neutrinos can be
generated in Model II by introducing small perturbations in the
lower right block of ${\cal M}$, i.e., small Majorana mass terms
for the neutral singlets $S_{i L}$, and this does not
significantly affect the mixings of heavy-to-light neutrinos.

In either model, the mass matrix ${\cal M}$ can always be
diagonalized by means of a congruent transformation involving a
unitary matrix $U$
\begin{eqnarray}
U {\cal M} U^{T} \Lambda^* &=& {\cal M}_d \ , \label{congruent}
\end{eqnarray}
where ${\cal M}_d = {\rm diag} (m_1, m_2, \ldots)$ is the
nonnegative diagonal mass matrix, and $\Lambda^*$ is an arbitrary
diagonal unitary matrix: $(\Lambda^*)_{ij} = \delta_{ij}
\lambda^*_i$ and $|\lambda_i|=1$. The $N_L+N_H$ mass eigenstates
$n_i$ are Majorana neutrinos, and are related to the interaction
eigenstates $\nu_{k L}$ and ${\widetilde \nu}_{j R}$ by the
matrices $U$ and $\Lambda^{\ast}$ as:
\begin{eqnarray}
\left(
\begin{array}{c}
\nu_L \\ {\widetilde \nu}_R^{ \ c}
\end{array}
\right)_a &=& \sum^{N_L+N_H}_{i=1}U^{*}_{ia} ~{n_{i L}} \quad
\Rightarrow \ \quad \left(
\begin{array}{c}
\nu_L^{ \ c} \\ {\widetilde \nu}_R
\end{array}
\right)_a
=
\sum^{N_L+N_H}_{i=1} U_{ia} ~\lambda^{*}_{i}~{n_{i R}} \ ,
\label{mixing}
\end{eqnarray}
The first $N_L$ eigenstates $n_i$ ($i=1,\ldots,N_L$) are the light
partners of the standard charged leptons. The other $N_H$
eigenstates are heavy. It can be checked from relations
(\ref{mixing}) that the relation $(n_{i L})^c = \lambda_i^* n_{i
R}$ holds, and thus $\lambda_i$ is recognized as the creation
phase factor \cite{Kayser:1984ge,Kayserb} of the Majorana neutrino
$n_i$.

In the mass basis, a $N_L \times (N_L\!+\!N_H)$-dimensional mixing
matrix $B$ for charged current interactions, and a $(N_L\!+\!N_H)
\times (N_L\!+\!N_H)$-dimensional matrix $C$ for neutral current
interactions can be introduced:
\begin{equation}
B_{l i}=U^{*}_{i l}, \qquad C_{ij}=\sum^{N_L}_{a=1}
U_{ia}U^{*}_{ja} \ . \label{BC}
\end{equation}

Accordingly, the leptonic weak interactions  are
(cf.~\cite{Pilaftsis:1991ug,CDKK2})
\begin{eqnarray}
{\cal L}_{l n W}(x) & = & \left( - \frac{g_w}{\sqrt{2}} \right)
W_{\mu}^-(x) \; \sum_{i=1}^{N_L} \sum_{j=1}^{N_L+N_H} B_{ij}
{\overline l}_i(x) \gamma^{\mu} P_L n_j(x)  + {\rm h.c.} \ ,
\label{lnW}
\\
{\cal L}_{l n G}(x) & = & \left(- \frac{g_w}{\sqrt{2} M_W} \right)
G^-(x) \sum_{i=1}^{N_L} \sum_{j=1}^{N_L+N_H} B_{ij} {\overline
l}_i(x) \left(m_{li}\, P_L - m_j P_R \right) n_j(x)  + {\rm h.c.}
\ , \label{lnG}
\\
{\cal L}_{n n Z}(x) & = & \left(- \frac{g_w}{4} \frac{M_Z}{M_W}
\right) Z^0_{\mu}(x) \; \sum_{i,j=1}^{N_L+N_H}{\overline
n}_i(x)\gamma^\mu \left(  C_{i j}P_L -  C_{i j}^\ast P_R \right)
n_j(x)
 + {\rm h.c.} \ ,
\label{nnZ}
\\
{\cal L}_{n n G}(x) & = & \frac{i g_w}{4 M_W} G^0(x)
\sum_{i,j=1}^{N_L+N_H} {\overline n}_i(x) \left\{ i\ (m_j -
m_i){\rm Im} C_{i j} + \gamma_5 ( m_j + m_i ) {\rm Re} C_{i j}
\right\} n_j(x) \ , \label{nnG}
\end{eqnarray}
where $G^-$ and $G^0$ are the Goldstone bosons that appear in a
general gauge. Here, the conventions of Itzykson and
Zuber\cite{IZ} are used; $P_{L} = (1 - \gamma_5)/2$, $P_{R} = (1 +
\gamma_5)/2$; $l_i$ is the $i$-th negatively charged lepton ($l_1
= e^-$, $l_2 = \mu^-$); $g_w$ is the $SU(2)_L$ coupling constant
($g_w^2 = 8 G_F M_W^2/\sqrt{2}$) and $n_j$ are the Majorana
neutrinos (mass eigenstates).

In all our numerical calculations we will just use two generations
({\it i.e.} $N_L = 2$), since a third generation does not make a
sizable change in our results, considering the uncertainties
involved.

\section{Transition amplitude for $\mu^+ e^-\to e^+ e^-$}
\label{sec:T}

The amplitude for this process is similar to those of Ref.\
\cite{IP} for lepton decays. Just as in that case, the amplitude
for $\mu^+ e^-\to e^+ e^-$
 can be calculated neglecting external momenta and
external masses inside the loop, provided the neutrinos in the
loop are heavy. Let us define the kinematics of the process as:
\begin{equation}
\mu^+(k_1)\ e^-(k_2)\ \to\  e^+(k_3)\  e^-(k_4). \label{scat-proc}
\end{equation}
The transition amplitude is a sum of photon- and Z- penguin
diagrams (Fig.~1) and box diagrams (Fig.~2). There are two types
of penguin diagrams in each case: those where the photon or Z is
in the t-channel and those where it is in the s-channel (see Fig.\
1). There is a relative minus sign between those diagrams due to
Fermi statistics. The contribution to the amplitude from the
photon penguins has two form factors, $F_\gamma^{e\mu}$ and
$G_\gamma^{e\mu}$ \cite{IP}:
\begin{eqnarray}
i{\cal T}_{\gamma} &=& i \frac{\alpha_w^2 \sin^2 \theta_w}{2 M_W^2} {\Bigg
\{}  \left[ {\overline v}_{\mu} (k_1) \left( F_{\gamma}^{e\mu}\
\big( \gamma^{\rho} - \frac{q^{\rho} {q \llap /}}{q^2} \big) P_L \
-\  i G_{\gamma}^{e\mu}\ \frac{ \sigma_{\rho \sigma}
q^{\sigma}}{q^2} (m_{e} P_R + m_{\mu} P_L) \right) u_{e}(k_2)
\right]\times \Big[ {\overline u}_{e}(k_4) \gamma_{\rho}
v_{e}(k_3) \Big] \nonumber\\ &&\qquad\qquad\qquad -\quad \Big[
v_{e}(k_3) \leftrightarrow u_{e} (k_2),\quad q \to q^{\prime}
\Big] {\Bigg \}} \ , \qquad (q = k_1 + k_2, \quad q^{\prime} = k_1
- k_3) \ .
\label{Tgamma1}
\end{eqnarray}
where $\alpha_w = g^2/(4 \pi)$,
$P_{R/L} = (1 \pm \gamma_5)/2$ are the chiral projectors, and
$\sigma_{\rho \sigma} = (i/2) [ \gamma_{\rho}, \gamma_{\sigma} ]$.
The term $q^\rho{q \llap /}$ vanishes when contracted to the vector
current of the $e^+ e^-$ pair.
Similarly, the Z penguins contain one form factor $F_Z^{e\mu}$:
\begin{eqnarray}
i{\cal T}_{Z} &=& i \frac{\alpha_w^2}{4 M_W^2}  {\Bigg
\{}F_Z^{e\mu}\ \Big[ {\overline v}_{\mu} (k_1) \gamma^{\rho} P_L
u_{e}(k_2) \Big] \times \Big[ {\overline u}_{e}(k_4) \gamma_{\rho}
\left( (1 - 2 \sin^2 \theta_w) P_L - 2 \sin^2 \theta_w P_R \right) v_e(k_3) \Big]
\nonumber\\ &&\qquad\qquad\qquad -\ \Big[ v_{e}(k_3)
\leftrightarrow u_{e} (k_2) \Big] {\Bigg \}} \ ,
\label{TZ1}
\end{eqnarray}
and the box diagrams also one form factor, $F_B^{e\mu e e}$:
\begin{eqnarray}
i{\cal T}_{{\rm Box}} \ = \  i \frac{\alpha_w^2}{4 M_W^2}
F_B^{e\mu e e} \Big[ {\overline v}_{\mu} (k_1) \gamma^{\rho} P_L
u_{e}(k_2) \Big]\times \Big[ {\overline u}_{e}(k_4) \gamma_{\rho}
P_L v_e(k_3) \Big] \ . \label{TBox1}
\end{eqnarray}

\noindent
\begin{figure}[htb]
\epsfig{file=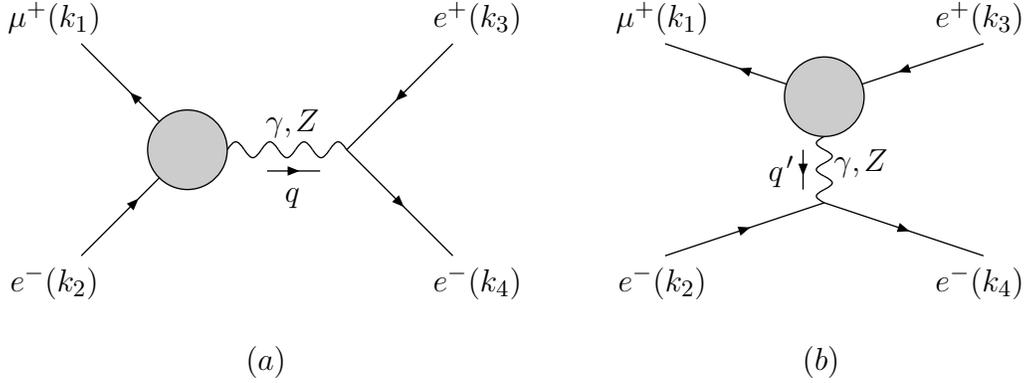}
\caption{
The one-loop penguin diagrams for $\mu^+(k_1) e^- (k_2)\to e^+
(k_3) e^- (k_4)$:  (a) s-channel penguin and (b) t-channel
penguin. The blob represents the loop involving $W, G$ bosons and
neutrinos. The diagram (b), {\it i.e.} t-channel, corresponds to
the contributions in the last lines of Eqs.~(\ref{Tgamma1}) and
(\ref{TZ1}).} \label{pengst}
\end{figure}

\begin{figure}[htb]
\epsfig{file=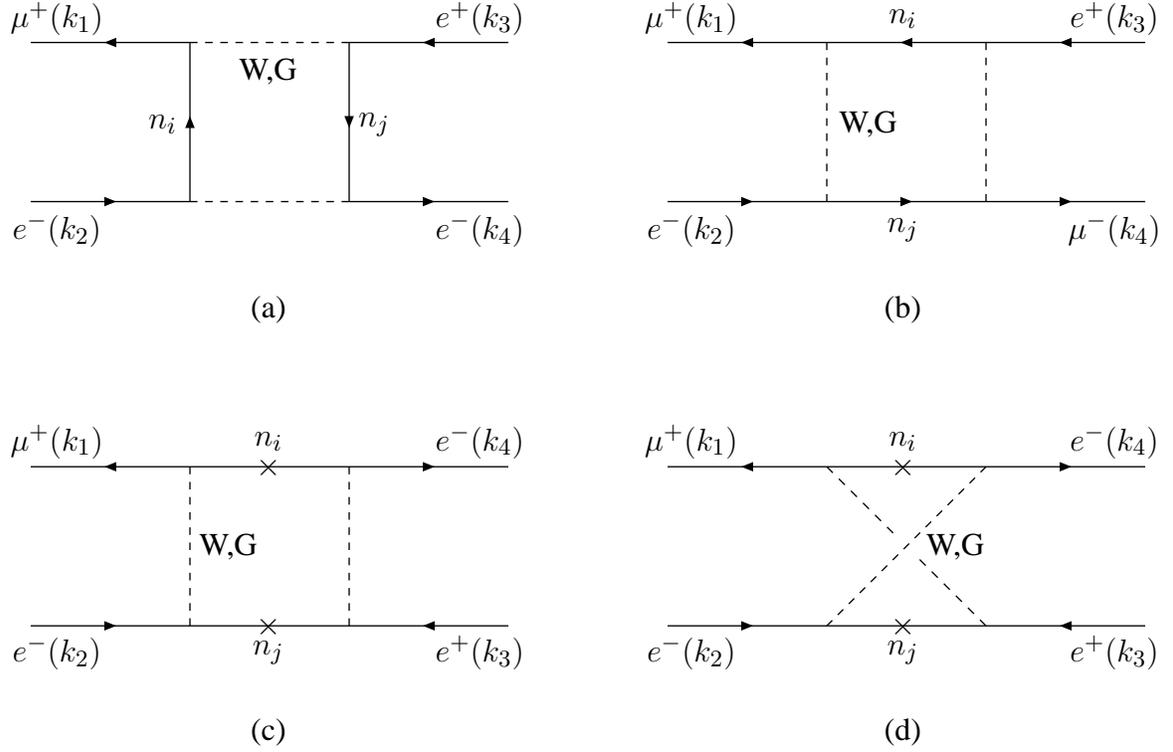}
\caption{
The box diagrams for $\mu^+(k_1) e^-(k_2) \to e^+(k_3) e^-(k_4)$.
The diagrams (b) and (d) are obtained from (a) and (c),
respectively, by the exchange $e^-(k_2) \leftrightarrow e^+(k_3)$
and by the corresponding change in the flow of time.}
\label{boxdec}
\end{figure}

The form factors $F_{\gamma}^{e \mu}$, $G_{\gamma}^{e \mu}$,
$F_Z^{e \mu}$ and $F_B^{e\mu e e}$ are
\begin{eqnarray}
F^{e \mu}_{\gamma} &=& \sum^{N_L+N_H}_{i=1}B^*_{e
i}B_{\mu i}\ F_{\gamma}(x_i) \approx
\sum_{I > N_L} B^*_{e I} B_{\mu I} F_{\gamma}(x_I)
\ , \label{Fg}
\\
G^{e \mu}_{\gamma} &=& \sum^{N_L+N_H}_{i=1}B^*_{e
i}B_{\mu i}\ G_{\gamma}(x_i) \approx
\sum_{I > N_L} B^*_{e I} B_{\mu I} G_{\gamma}(x_I)
\ , \label{Gg}
\\
F^{e \mu}_Z &=& \sum^{N_L+N_H}_{i,j=1}B^*_{e i}B_{\mu j}
\Big[\delta_{ij}\ F_Z(x_i)+C_{ij}\ H_Z(x_i,x_j) +C^*_{ij}\ G_Z(x_i,x_j)\Big] \label{FZa}
\\
& \approx &
\sum_{I > N_L} B^*_{e I} B_{\mu I} F_Z(x_I) +
\sum_{I,J > N_L} B^*_{e I} B_{\mu J} \left[ C_{I J} H_Z(x_I,x_J) +
C^*_{I J} \left( G_Z(x_I,x_J) - G_Z(x_I,0) - G_Z(0,x_J) \right) \right]
\nonumber\\
&&+ 2 \sum_{I > N_L} B^*_{e I} B_{\mu I} G_Z(x_I,0) \ ,
\label{FZb}
\\
F^{e \mu e e}_{B} &=& \sum^{N_L+N_H}_{i,j=1} \Big[
2 | B_{e i} |^2 B^*_{e j} B_{\mu j} \ F_{\rm Box}(x_i,x_j) -
( B^*_{e i} )^2  B_{e j} B_{\mu j} \lambda_j \lambda^*_i \
G_{\rm Box}(x_i,x_j) \Big]
\label{FBox1a}
\\
& \approx & 2 \sum_{I, J > N_L}
| B_{e I} |^2 B^*_{e J} B_{\mu J} \left[
F_{\rm Box}(x_I, x_J) + F_{\rm Box}(0,0) - F_{\rm Box}(x_I,0)
- F_{\rm Box}(0,x_J) \right]
\nonumber\\
&&
- \sum_{I, J > N_L}
( B^*_{e I} )^2  B_{e J} B_{\mu J} \lambda_J \lambda^*_I \
G_{\rm Box}(x_I,x_J)
+ 2 \sum_{J > N_L} B^*_{e J} B_{\mu J} \left(
F_{\rm Box}(0,x_J) - F_{\rm Box}(0,0) \right) \ ,
\label{FBox1b}
\end{eqnarray}
where $x_i = m_i^2/M_W^2$ ($m_i$ is the mass of the Majorana
neutrino of type $i$ inside the loop), and the functions
$F_{\gamma}(x)$, $G_{\gamma}(x)$, $F_Z(x)$, $H_Z(x,y)$,
$G_Z(x,y)$, $F_{\rm Box}(x,y)$ and $G_{\rm Box}(x,y)$ appear from
the loop integrals \cite{IP}, in the approximation where external
masses and momenta are neglected. The explicit expressions of
these functions are given in Ref.~\cite{CDKK2}. $F_{\rm Box}$
corresponds to the diagrams of Figs.~\ref{boxdec}(a) and (b), and
$G_{\rm Box}$ to those of Figs.~\ref{boxdec}(c) and (d) which
exist only for Majorana neutrinos. Eqs.~(\ref{Fg})-(\ref{FBox1a})
involve summation over all 2-component neutrinos, where $N_L$ is
the number of standard (light) ones and $N_H$ the number of extra
(heavy) ones ($=\!N_L, 2 N_L$ in Models I, II, respectively). All
these expressions are identical to those obtained in
Ref.~\cite{IP} for neutrinoless lepton decays, except for the sign
in front of the term proportional to $G_{\rm Box}(x_i,x_j)$ -- for
details on the latter point, cf.~Ref.~\cite{CDKK2}. Furthermore,
in Eqs.~(\ref{Fg})-(\ref{FBox1b}) the expressions in a modified
form are included which involve only the light-to-heavy mixing
elements $B_{\ell I}$ ($\ell \leq N_L; I > N_L$); they are
obtained by taking into account the unitarity of the $U$ matrix,
and the fact that the light neutrino masses are $m_{\ell} \ll
M_W$.

The parameters $\lambda_i$ are the creation phase
factors\cite{Kayserb} of the Majorana neutrinos $n_i$ and $n_j$,
respectively ($n_j^c = \lambda_j^{\ast} n_j$), which appear in the
matrix $\Lambda$ of Eq.~(\ref{congruent}). In the case of no CP
violation and real mass matrix ${\cal M}$, we can choose the
unitary matrix $U$ as a real orthogonal matrix which diagonalizes
the symmetric and real matrix ${\cal M}$; then some of the
$\lambda_j$'s are $+1$ and others are $-1$ \cite{CDKK2}.
Nonetheless we can choose all $\lambda_j = +1$, but then in
general the unitary matrix $U$ will not be real.

The total amplitude for $\mu^+ e^-\to e^+ e^-$ is then the sum of
(\ref{Tgamma1}), (\ref{TZ1}) and (\ref{TBox1}).

\section{The decay of muonium into $e^- e^+$}
\label{sec:Mtoeeform}

The calculation of the decay width of muonium $M$ into $e^-e^+$ is
determined by the LFV transition $\mu^+ e^-\to e^+ e^-$ at low
momentum and by the muonium bound state wavefunction. The formula
for the decay width of muonium $M$ into a final state $f$ is
analogous to the expression for positronium decay:
\begin{equation}
d\Gamma(M \to f) = \frac{|\psi(0)|^2}{4\mu M}\ \overline{|\langle
f|{\cal T}|\mu,e\rangle|^2}\  dLips. \label{GMSM1}
\end{equation}
Here $\mu$ and $M$ are the reduced mass and total mass of the
bound state, respectively; $\psi(0)$ is the spatial wavefunction
of the bound state at ${\bf x}=0$; the bar over the matrix element
means average over the initial spin components; and $dLips$ is the
usual {\sl Lorentz invariant phase space} of the final state,
$dLips = \delta^4 (p_f - p_i)\ d^3
p_3 \ d^4 p_4 /(16\pi^2 E_3 E_4)$.

The width can also be expressed in terms of the cross section of
the constituents at low momentum, $\sigma(\mu^+ e^-\to f)$, since
the latter involves the same transition element and phase space
integral:
\begin{eqnarray}
\Gamma(M\to f) = |\psi(0)|^2 \frac{{\cal S}_{12}}{2\mu M}\
\sigma(\mu^+ e^-\to f).
 \label{decay-cross}
 \end{eqnarray}
 Here ${\mathcal S}_{12}\equiv 2\ E_{CM}\ p_{CM}$ is the flux
factor. At low momenta, the factor ${\cal S}_{12}/2\mu M$ reduces
to the relative velocity between the constituents, $v_{\rm rel}$.
Since muonium is a rather nonrelativistic Coulomb bound state, the
wavefunction is calculable:
\begin{eqnarray}
|\psi(0)|^2 = \frac{ \alpha_{em}^3\ \mu^3}{\pi}\quad =
\frac{ \alpha_{em}^3}{\pi}
\frac{m_e^3}{(1+m_e/m_\mu)^3}
\label{wfsq}
\end{eqnarray}
and is independent of the final state $f$ into which the
constituents disintegrate. On the other hand, the disintegration
process itself is totally contained in the cross section
$\sigma(\mu^+ e^-\to f)$. Using the two previous expressions, the
decay $M \to e^- e^+$ results in:
\begin{equation}
\Gamma(M \to e^- e^+) = \left[ \frac{\alpha_{em}^3}{\pi} \frac{
m_e^3}{(1+m_e/m_\mu)^3} \right]v_{\rm rel}\ \sigma(e^- \mu^+ \to
e^- e^+) {\Bigg|}_{v_{\rm rel} \to 0} \ . \label{GMee}
\end{equation}
Therefore, the calculation of this decay rate reduces to the
calculation of the scattering cross section $\sigma(e^- \mu^+ \to
e^- e^+)$ at low (nonrelativistic) momentum. We remark that the
factor $v_{\rm rel}$ in Eq.~(\ref{GMee}) will cancel the
corresponding factor $1/v_{\rm rel}$ contained in the cross
section. Now, let us find the decay rate within Models I and II
described in the previous section.

Although our expressions for the amplitude and form factors are
the same as those of Ref.\ \cite{IP} for lepton decays, the matrix
element for muonium decay becomes a different combination of form
factors, because the kinematic regime in muonium decay is quite
different than that in free lepton decay: in muonium, we should
take the initial 3-momenta as non-relativistic (or vanishing,
since the amplitudes are smooth functions of momenta). The result
for the total amplitude squared, summed over final spins and
averaged over initial spins, is then:
\begin{eqnarray}
\overline{|{\cal T}_{\rm total}|^2} =
\frac{\alpha_w^4}{16 M_W^4}
\bigg\{
&& \left( m_e\,m_\mu^{3}+2\,{m_e}^{2}m_\mu^{2}
+{m_e}^{3}m_\mu \right) {\left| 2F_Z^{e\mu} + F_B^{e\mu ee}\right|}^{2}
\nonumber\\
&+& 4 \sin^2 \theta_w\left( 2\,m_e\,m_\mu^{3}
+3\,{m_e}^{2}m_\mu^{2}+3\,{m_e}^{3}m_\mu \right)\
{\rm Re} \left[(2F_Z^{e\mu}+F_B^{e\mu e e})
(F_\gamma^{e\mu} -F_Z^{e\mu})^*\right]\nonumber\\
&+& 12 \sin^2 \theta_w
\left( m_e\,m_\mu^{3}+ 2\,{m_e}^{2}m_\mu^{2}+{m_e}^{3}m_\mu \right)
\ {\rm Re} \left[(2F_Z^{e\mu}+F_B^{e\mu e e})
G_\gamma^{e\mu\ *}\right] \nonumber\\
 &+& 4 \sin^4\theta_w \left( 7\,m_e\,{m_\mu}^{3}+12\,{m_e}^{2}m_\mu^{2}
 +9\,{m_e}^{3}m_\mu \right) {\left|F_\gamma^{e\mu}-F_Z^{e\mu}\right|}^{2}\nonumber\\
&+& 4 \sin^4\theta_w \left( -2\,m_\mu^{4}
+12\,m_e\,m_\mu^{3}+36\,{m_e}^{2}m_\mu^{2}+18\,{m_e}^{3}m_\mu \right)
\ {\rm Re} \left[(F_\gamma^{e\mu} - F_Z^{e\mu}) G_\gamma^{e\mu\ *}
\right]\nonumber\\
&+& 4 \sin^4\theta_w \left( {\frac {m_\mu^{5}}{m_e}}+2\,m_\mu^{4}
+8\,m_e\,{m_\mu}^{3}+24\,{m_e}^{2}m_\mu^{2}
+9\,{m_e}^{3}m_\mu \right) \left|{G_\gamma^{e\mu}}\right|^{2}\bigg\}
\ .
\label{T2}
\end{eqnarray}
From this expression and Eq.~(\ref{GMee}) one obtains the muonium
decay rate, after integration over phase space. Since the
amplitude is independent of the kinematics in the non-relativistic
limit, the integration is simply $\int dLips = (1/8\pi)\sqrt{1-4
m_e^2/(m_\mu + m_e)^2}$, and we have
\begin{eqnarray}
\Gamma(M \to e^+ e^-) & = & \frac{\alpha_{em}^3 m_e}{32 \pi^2}
\frac{m_e}{m_{\mu}} \frac{m_{\mu}^3}{(m_{\mu} + m_e)^3}
\sqrt{\left[ 1 - 4 \frac{m_e^2}{(m_\mu + m_e)^2} \right]}
\; \overline{|{\cal T}_{\rm total}|^2}
\label{M-rate1a}
\\
&=& \frac{\alpha_{em}^3 m_e}{32 \pi^2} \frac{m_e}{m_{\mu}}
\left[ 1 - 3 \frac{m_e}{m_{\mu}} +
{\cal O} \left( \frac{m_e^2}{m_{\mu}^2} \right) \right]
\overline{|{\cal T}_{total}|^2} \ ,
\label{M-rate1b}
\end{eqnarray}
where $\overline{|{\cal T}_{total}|^2}$ is given in
Eq.~(\ref{T2}). Now, comparing Eq.\ (\ref{M-rate1a}) with the
tree-level expression for the muon width $\Gamma(\mu\to
all)^{(0)}$ $= G_F^2 m_\mu^5/(192\pi^3)$, which is comparable to
the {\sl muonium} width, we obtain the branching ratio:
\begin{eqnarray}
Br(M\to e^+ e^-) &\equiv& \frac{\Gamma(M\to e^+
e^-)}{\Gamma(\mu\to all)^{(0)}}\nonumber\\ &=& \frac{3}{\pi}
\alpha_{em}^5 \left( \frac{m_e}{m_{\mu}} \right)
\frac{m_{\mu}^3}{(m_{\mu} + m_e)^3} \sqrt{\left[ 1 - 4
\frac{m_e^2}{(m_\mu + m_e)^2} \right]} \frac{1}{\sin^4 \theta_w}
B_{\rm red.} \approx 1.77 \cdot 10^{-12} \ B_{ \rm red.} \ ,
\label{M-rate2}
\end{eqnarray}
where we defined a reduced branching ratio $B_{\rm red.}$:
\begin{equation}
B_{\rm red.} = \frac{4}{\alpha_W^4} \frac{M_W^4 m_e}{m_{\mu}^5}
\overline{|{\cal T}_{total}|^2} \ , \label{Gred}
\end{equation}
 with $\overline{|{\cal T}_{total}|^2}$ given by
(\ref{T2}). Provided the neutrino masses are below a few TeV, a
rather good approximation for $B_{\rm red.}$ is obtained by
expanding Eq.~(\ref{T2}) to second order in $(m_e/m_{\mu})$:
\begin{eqnarray}
B_{ \rm red.}  & = & {\bigg\{} \sin^4 \theta_w |G_{\gamma}|^2 - 2
\left( \frac{m_e}{m_{\mu}} \right) \sin^4 \theta_w  {\rm Re}
\left[ (F_{\gamma} - F_Z) G^*_{\gamma} - |G_{\gamma}|^2 \right]
\nonumber\\ &&+\ \frac{1}{4} \left( \frac{m_e}{m_{\mu}} \right)^2
{\Big [} |F_B + 2 F_Z + 4 \sin^2 \theta_w (F_{\gamma} - F_Z) + 6
\sin^2 \theta_w G_{\gamma} |^2 \nonumber\\ && +\ 4 \sin^4 \theta_w
\left( 3 |F_{\gamma}|^2 - |G_{\gamma}|^2 + 3 |F_Z|^2 - 6 {\rm
Re}(F_Z F^*_{\gamma}) \right) {\Big ]} + {\cal O}\left(
\frac{m_e^3}{m_{\mu}^3} \right) {\bigg \}} \ . \label{Gred2}
\end{eqnarray}
For simplicity of notation we have omitted the superscripts $e
\mu$ and $e \mu ee$ in the form factors. We see that the box
diagrams ({\it i.e.}, the form factors $F_{\rm Box}$) have in
general only a small effect on this process. This contrasts with
muonium to antimuonium conversion, where the box diagrams are the
only contribution \cite{CDKK2}. This is a bit disappointing, since
it is only in the box diagrams where the true Majorana character
of neutrinos appear, as seen in Figs.\ 2(c) and (d). The rest of
the diagrams are sensitive to the neutrino masses and mixings,
irrespective of their Dirac or Majorana character. In the next
section we will see, however, that the free $\mu^+ e^-$ scattering
at high energy depends on a different combination of form factors,
where the box diagrams do enter. In this sense, $\mu^+ e^-$
scattering at high energy is complementary to muonium decay, when
testing this LFV process.
\begin{figure}[htb]
\setlength{\unitlength}{1.cm}
\begin{center}
\epsfig{file=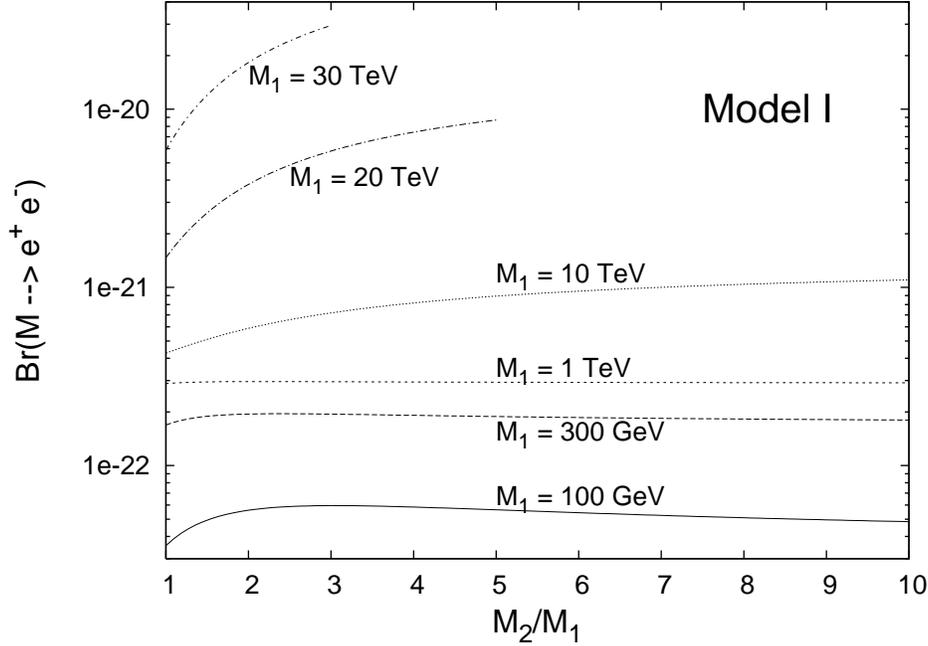}
\end{center}
\vspace{0.0cm} \caption{\footnotesize The branching ratio
$\Gamma(M \to e^+e^-)/\Gamma(\mu \to all)^{(0)}$,
Eq.~(\ref{M-rate2}) in Model I, for the (approximately) optimal
choice [cf.~Eq.~(\ref{MIopt})] for the Dirac mass parameters, as a
function of the neutrino mass ratio $M_2/M_1$ ($\geq 1$), and for
different $M_1= 100$ GeV, $300$ GeV, $1$ TeV, $10$ TeV, $20$ TeV and
$30$ TeV, as indicated. The last two curves are shown only up to
$M_2/M_1 \approx 5$ and $3$, respectively,
due to the perturbative unitarity bound \cite{IP}.}
\label{MdecI}
\end{figure}

In order to obtain numerical results, the parameter space of the
models must first be reasonably narrowed down. First the
parameters appearing in the Dirac matrix $m_D$, Eq.~(\ref{Dm1}),
are chosen in such a way as to fulfill, at fixed Majorana masses
$M_1$ and $M_2$, all the experimental requirements, and at the
same time to approach maximal squared amplitudes (e.g.\
$|G_{\gamma}|^2$ etc.) appearing in the decay width of Eqs.\
(\ref{M-rate2}) and (\ref{Gred}) (see Refs.~\cite{CDKK,CDKK2} for
details), in order to find the most favorable case to observe such
rare processes. This optimal condition fixes the model parameters
as follows:
\begin{eqnarray}
{\rm Model \ I:} \quad a &= &c = \frac{M_1 s_L}{\sqrt{1 +
M_1/M_2}} \ , \nonumber\\ b &=& - d = a \sqrt{M_2/M_1} \ , \quad
\delta_1 = \delta_2 = \pi/2 \ , \label{MIopt}
\\
{\rm Model \ II:} \quad
a & = & \frac{M_2 s_L^{\nu_e}}
{\sqrt{(M_2/M_1)^2 + (s_L^{\nu_{\mu}}/s_L^{\nu_e})^2 }} \ ,
\nonumber\\
b & = & c = a \times (s_L^{\nu_{\mu}}/s_L^{\nu_e}) \ ,
\quad
d = a \times (s_L^{\nu_{\mu}}/s_L^{\nu_e})^2 \ ,
\quad
\delta_1 = \delta_2 = 0 \ .
\label{MIIopt}
\end{eqnarray}
Here, $(s_L^{\nu_{\ell}})^2 \equiv \sum_{h} |B_{\ell h}|^2$ ($N_L
< h \leq N_L+N_H$; $\ell = e, \mu$) are the light-to-heavy
mixings, and $s_L$ in (\ref{MIopt}) means the
common value $s_L^{\nu_{e}} =  s_L^{\nu_{\mu}}$.
 The squares $|A_{\rm p.}|^2$ of the penguin diagrams
$A_{\rm p.}$, for the parameter choices (\ref{MIopt}) and
(\ref{MIIopt}), are approximately proportional to
$(s_L^{\nu_{\mu}})^2 (s_L^{\nu_e})^2$, as shown
in Appendix of Ref.~\cite{CDKK} on the basis
of unitarity of matrix $U$. The quantity
$Br(M \to e^+e^-)$ is dominated by the penguin
diagrams, as seen earlier in this Section. Further,
the other quantities considered later on in this
work (cross sections for $e^- \mu^+ \to e^-e^+$
and for $e^- \mu^+ \to WW$), are also approximately
proportional to $(s_L^{\nu_{\mu}})^2 (s_L^{\nu_e})^2$.\footnote{
For the box diagrams there can be substantial deviations
from such proportionality, especially when matrix
$U$ is not real.}
While each of the two mixings separately is
restricted by inequalities \cite{Nardi}
\begin{equation}
(s_L^{\nu_e})^2 \leq 0.005 \ , \qquad
 (s_L^{\nu_{\mu}})^2 \leq 0.002 \ ,
\label{sLmax1}
\end{equation}
there is a strong restriction on the product
of the two mixings: $s_L^{\nu_e} s_L^{\nu_{\mu}} < 0.00012$.
The latter restriction is obtained in the two models,
with parameter choice (\ref{MIopt}) and (\ref{MIIopt}),
when the models are required to
give predictions for the branching ratio
$Br(\mu^- \to e^- \gamma)$ which respect the
experimental upper bound \cite{PDG2006}:
\begin{equation}
Br(\mu^- \to e^- \gamma) < 1.2 \times 10^{-11} \ .
\label{muegamma}
\end{equation}
More specifically, the aforementioned upper bound for
$s_L^{\nu_e} s_L^{\nu_{\mu}}$
is obtained from restriction (\ref{muegamma}) because
\begin{equation}
Br(\mu^- \to e^- \gamma) = \alpha_{em} \frac{3}{2 \pi} |G_{\gamma}^{e \mu}|^2
\ ,
\label{muegammath}
\end{equation}
and it can be shown that
$|G_{\gamma}^{e \mu}|^2 \leq (1/4) (s_L^{\nu_{\mu}})^2 (s_L^{\nu_e})^2$,
where approximate equality is obtained when the parameters are
chosen according to Eqs.~(\ref{MIopt}) and (\ref{MIIopt}) in models
I and II, respectively.
We will choose $s_L^{\nu_e} = s_L^{\nu_{\mu}} \equiv s_L$. Therefore,
we will take in (\ref{MIopt}) and (\ref{MIIopt}) the
saturated values
\begin{equation}
(s_L)^2 = (s_L^{\nu_{e}})^2 = (s_L^{\nu_{\mu}})^2 = 1.2 \times 10^{-4} \ .
\label{sLmax}
\end{equation}
For simplicity, we restricted Model II to cases with no
CP violation ($\delta_1 = \delta_2 = 0$). The heavy neutrino
masses $M_1$ and $M_2$ are restricted to be above $100$ GeV, and
we take the convention $M_1 < M_2$. There is also an upper bound
for $M_1$ and $M_2$ in the form $M_h^2 \sum_{\ell} |B_{\ell h}|^2
< 2 M^2_W/\alpha_W$, with $\ell = e, \mu$ \cite{IP}, due to the
breakdown of perturbative expansion to one loop above this bound.
In Model I, this bound, in conjunction with
restrictions (\ref{sLmax}), translates into a restriction which is
approximately of the form $M_1 \cdot M_2 \alt (50$ TeV$)^2$, while
for Model II the bound allows any value of $M_2$ ($\geq M_1$) as
long as $M_1 \alt 60$ TeV.

\begin{figure}[htb]
\setlength{\unitlength}{1.cm}
\begin{center}
\epsfig{file=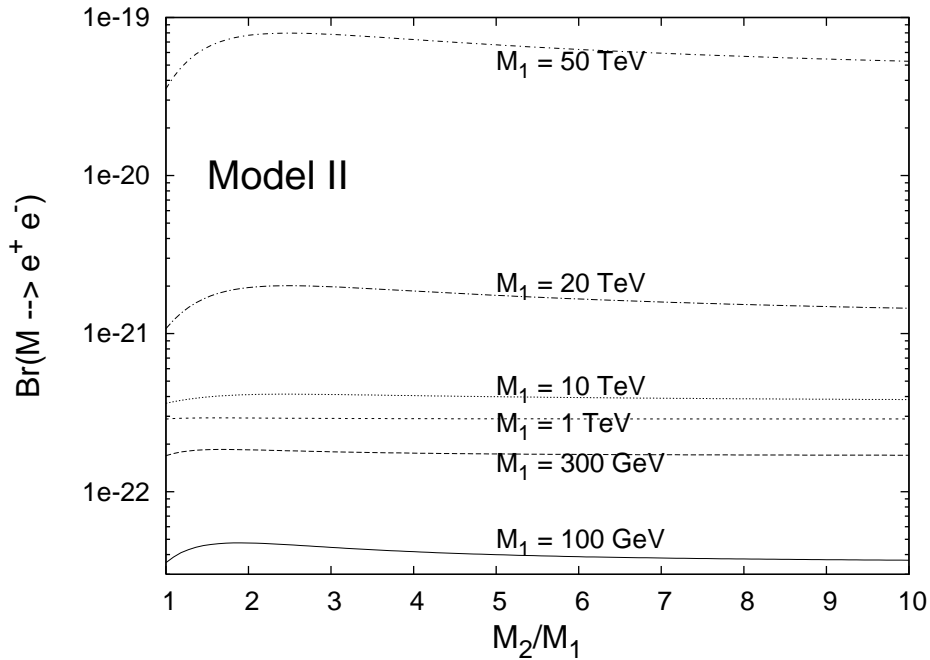}
\end{center}
\vspace{0.0cm}
\caption{\footnotesize
Same as in Fig.~\ref{MdecI}, but this time for Model II,
for the (approximately) optimal
choice (\ref{MIIopt}) for the Dirac mass parameters.
The curves, including the ones with $M_1 = 20$ and $50$ TeV,
are in this model allowed by the PUB.}
\label{MdecII}
\end{figure}

The numerical results for the branching ratio $Br(M\to e^+ e^-)$
defined in Eq.~(\ref{M-rate2}) are given in Fig.~\ref{MdecI} and
Fig.~\ref{MdecII}, for Model I and Model II, respectively.
We see that the branching ratios are extremely small, mainly
because of the factor $\sim 10^{-12}$ appearing in
Eq.~(\ref{M-rate2}). Two major sources for this suppression are a
factor $\alpha_{em}^3$ coming from the muonium wavefunction and a
factor $\alpha_{em}^2$ from the one-loop character of the squared
amplitude. Another source of suppression are the small
upper bounds (\ref{sLmax}) on the mixing-parameters.
From Figs.~\ref{MdecI} and \ref{MdecII} we see that the
branching ratios in in both models canot surpass
$10^{-19}$. These values are so small
that they seem out of the reach of experiments in the foreseeable
future.

\section{$\mu^+ e^- \to e^+ e^-$ collisions at high energy}
\label{sec:cs}

\begin{figure}[htb]
\begin{minipage}[b]{.49\linewidth}
 \centering\epsfig{file=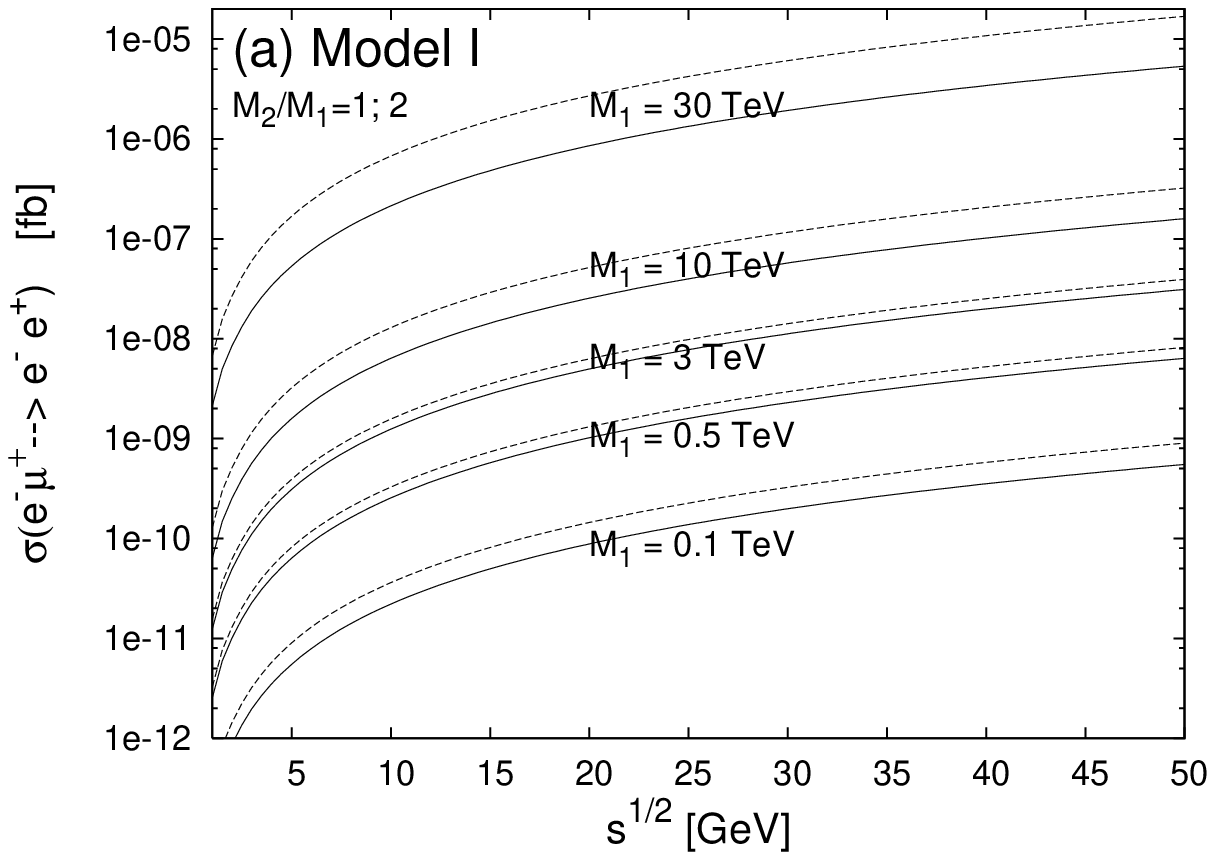,width=\linewidth}
\end{minipage}
\begin{minipage}[b]{.49\linewidth}
 \centering\epsfig{file=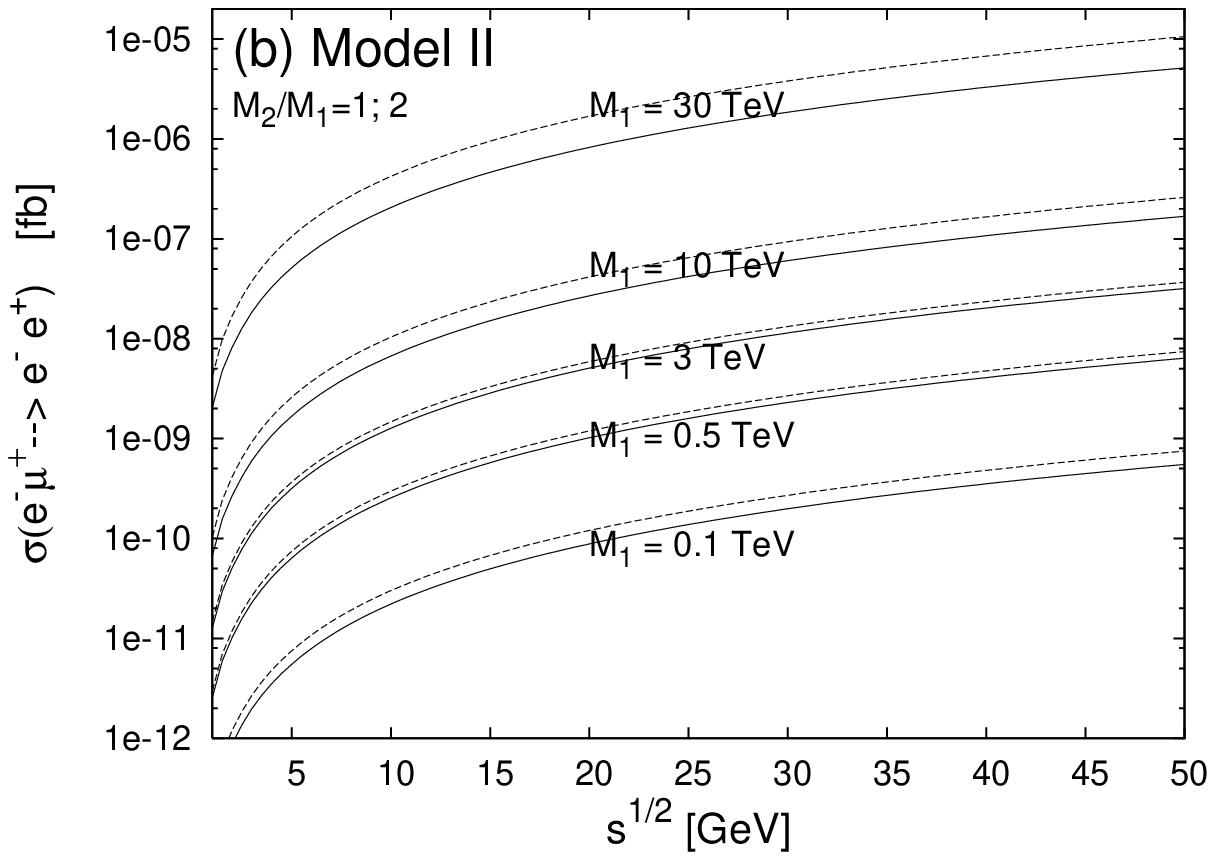,width=\linewidth}
\end{minipage}
\caption{\footnotesize The cross section for the LFV scattering
process $e^- \mu^+ \to e^- e^+$ as a function of CMS energy
$\sqrt{s}$ in (a) Model I, and (b) Model II. In all cases, each
pair of curves corresponds to a given value of $M_1$ and to the
values of $M_2/M_1 = 1$ (solid line) and $M_2/M_1 = 2$ (dotted
line).} \label{emuee}
\end{figure}

In addition to muonium decay, one can study the free scattering
process $\mu^+ e^-\to e^+ e^-$ at higher energies. The cross
section in the center-of-mass system (CMS) for this process at
energies well above the muon mass follows from the same amplitudes
(\ref{Tgamma1}), (\ref{TZ1}) and (\ref{TBox1}), but the form
factors enter in a quite different combination because of the
different kinematic regime. Indeed, for CM energies $\sqrt{s}$ in
the range $m_\mu \ll \sqrt{s} \ll M_W$, the cross section is:
\begin{eqnarray}
\sigma ( e^- \mu^+ &\to& e^- e^+; \; s)\\
&=&\frac{1}{12 \pi} \left( \frac{\alpha_w^2}{4 M_W^2} \right)^2
s \left[ {\big |} (1 - 2 \sin^2 \theta_w) F_Z^{e \mu} +
2 \sin^2 \theta_w F_{\gamma}^{e \mu}
+ \frac{1}{2} F_{\rm Box}^{e \mu e e } {\big |}^2 +
4 \sin^4\theta_w {\big |} F_{\gamma}^{e \mu} - F_{Z}^{e \mu} {\big |}^2
\right] .
\label{sigemuee}
\nonumber
\end{eqnarray}
The numerical results, for Models I and II at various chosen heavy
neutrino masses $M_1$ and $M_2$, are given in Figs.~\ref{emuee}
(a) and (b), respectively.
The elements of the Dirac mass $m_D$ are again chosen in the
optimized forms (\ref{MIopt}) and (\ref{MIIopt}), respectively,
and the mixing parameters have the upper bound values
(\ref{sLmax}).

As shown in Figs.~\ref{emuee}, the cross sections in both models
are of the same order of magnitude, the ones in Model I being
slightly larger than the corresponding ones in Model I.
These cross sections are
however very small. If one expects integrated luminosities not
much larger than a few fb$^{-1}$ at an hypothetical muon collider,
then an experimental test of this process is not likely to be
viable.

\section{Unsuppressed scattering channel $e \mu \to W W$}
\label{sec:emuWW}

To complement our analysis of the LFV scattering process $\mu^+
e^-\to e^+ e^-$, we consider the case of open $W$ production.

\begin{figure}[htb]
 \centering\epsfig{file=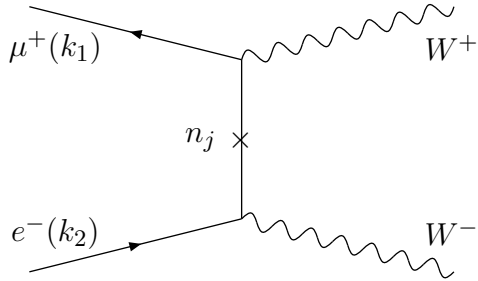}
\caption{\footnotesize The tree-level process $e^- \mu^+ \to W^-
W^+$ via exchange of a Majorana neutrino.} \label{figmueWW}
\end{figure}

When we consider  $e^- \mu^+$ scattering at CMS energies above the
$W^+ W^-$ production threshold, the unsuppressed tree-level
channel $e^- \mu^+ \to W^- W^+$ opens up. This process is mediated
by a neutrino exchange, as depicted in Fig.~\ref{figmueWW}.  Here
we will limit ourselves to estimate the cross section of the
process. However, it is understood that the experimental
observation of this process is hindered not just by the smallness
of the cross section, but also by the rejection of background
events that contain neutrinos in the final state.

The amplitude ${\cal T}$ for $\mu^+ e^- \to W^+ W^-$ is
\begin{equation}
{\cal T} = - i \frac{g_w^2}{2} B_{\mu j} B_{e j}^{\ast} \frac{1}{
\left[ (p_1 - k_1)^2 - m_j^2 + i \varepsilon \right] } \left[
{\overline v}(k_1) {\epsilon \llap /}(p_1)^{\ast} ( {p \llap /}_1
- {k \llap /}_1 )  {\epsilon \llap /}(p_2)^{\ast} P_L u(k_2)
\right] \ , \label{TmueWW}
\end{equation}
where $\epsilon$ is the W-polarization vector. Consequently the
total cross section for the process is:
\begin{eqnarray}
\sigma(e^- \mu^+ \to WW; \; s) &=& \frac{1}{64 \pi}
\left( \frac{g_w^2}{2} \right)^2 \frac{1}{s^2}
\sum_{i,j = 1}^{N_L+N_H} \left( B_{e j}^{\ast} B_{\mu j}
B_{e i} B_{\mu i}^{\ast} \right) A_{i j}
\label{sigemuWW1}
\\
& = &
\frac{1}{64 \pi}
\left( \frac{g_w^2}{2} \right)^2 \frac{1}{s^2}
\sum_{I,J > N_L} \left( B_{e J}^{\ast} B_{\mu J}
B_{e I} B_{\mu I}^{\ast} \right) \left( A_{I J} + A_{\ell \ell}
- A_{\ell J} - A_{I \ell} \right)
\label{sigemuWW2}
\ ,
\end{eqnarray}
where the explicit expressions of the coefficients $A_{i
j}(m_i^2,m_j^2,M_W^2,s)$ are written in Appendix. To go from
Eq.~(\ref{sigemuWW1}) to Eq.~(\ref{sigemuWW2}) we used unitarity
of the mixing matrix (the index $\ell$ in $A_{ij}$ denotes a light
neutrino state, whose mass we can neglect). The Dirac mass matrix
$m_D$ for models I and II is again chosen in the forms
(\ref{MIopt}) and (\ref{MIIopt}), respectively, and the
mixing-parameter values are those of Eq.~(\ref{sLmax}).

\begin{figure}[htb]
\begin{minipage}[b]{.49\linewidth}
 \centering\epsfig{file=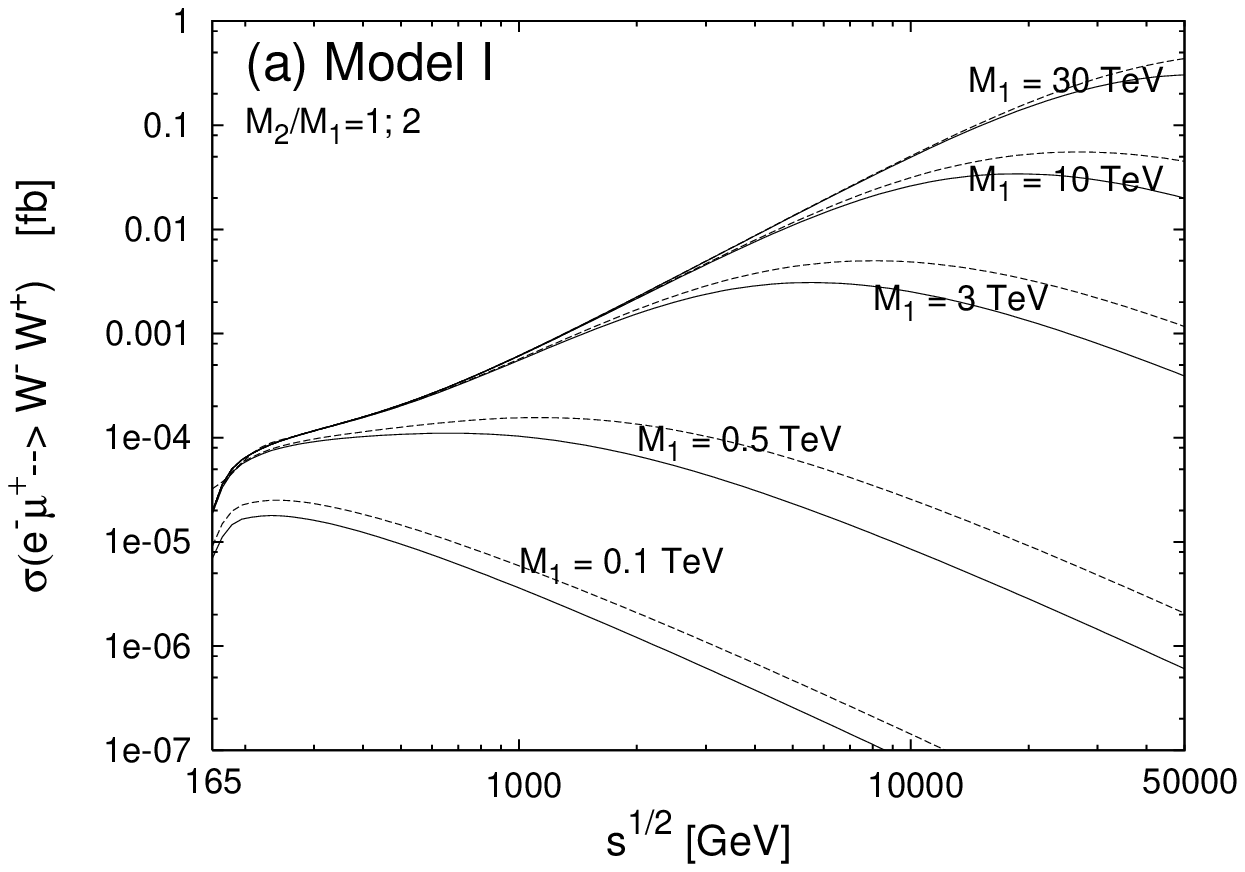,width=\linewidth}
\end{minipage}
\begin{minipage}[b]{.49\linewidth}
 \centering\epsfig{file=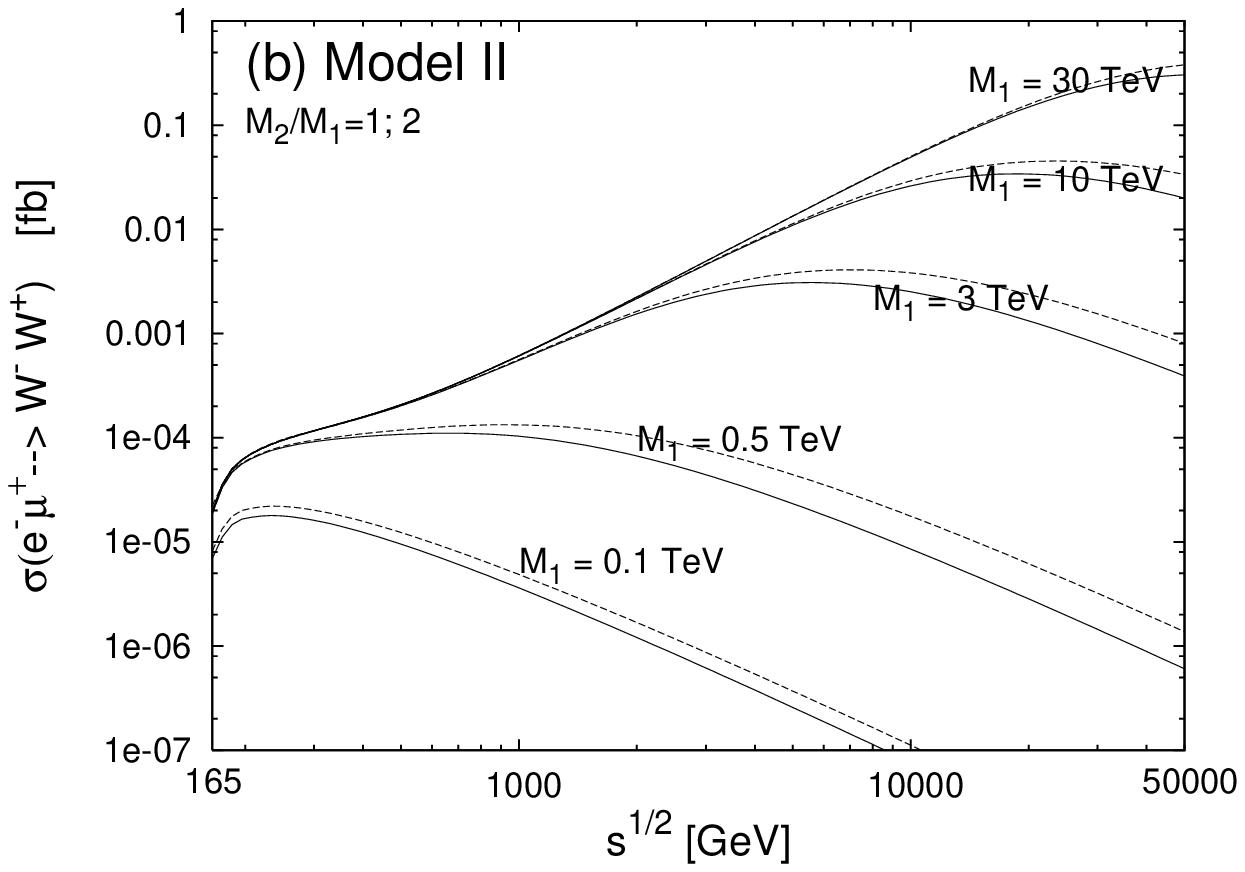,width=\linewidth}
\end{minipage}
\caption{\footnotesize The cross section for the LFV scattering
process $e^- \mu^+ \to W^- W^+$ as a function of CMS energy
$\sqrt{s}$ in (a) Model I, and (b) Model II. Each pair of curves
corresponds to the indicated value of $M_1$, and $M_2/M_1=1$
(solid line) or $2$ (dotted line).} \label{emuWW}
\end{figure}

The numerical results for Models I and II, and various chosen
heavy Majorana neutrino masses $M_1$ and $M_2$, are given in
Figs.~\ref{emuWW} (a) and (b), respectively.

We can see that the open $W^+W^-$ production cross section at
$\sqrt{s}\sim 200$ GeV is between $10^{-5}$ and $10^{-4}$ fb,
which is in general several orders of magnitude larger than the cross
section for $e^+e^-$ production for CMS energies below the
$W^+W^-$ threshold,\footnote{
If $M_1$ and $M_2$ are very large ($> 10$ TeV), the two
cross sections at $\sqrt{s} \sim 200$ GeV are comparable,
$\sim 10^{-4}$-$10^{-5}$ fb.}
but it is still very small. As a reference,
the point cross section
$\sigma(e^+e^-\to\gamma^\ast\to\mu^+\mu^-)$ at $\sqrt{s} = 200$
GeV is close to 2 pb, which is 7 orders of magnitude larger.

\section{Conclusions}
\label{conclusions} In the present work we have studied the
possibility to observe lepton flavor violating effects in $\mu^+
e^-$ annihilation processes induced by heavy neutrino scenarios.
We use two scenarios: one that is a plain seesaw model that
violates both lepton flavor and lepton number (Model I) and
another that has a richer spectrum of neutral fermions and
violates lepton flavor while conserving lepton number (Model II).
In a previous work\cite{CDKK2} we have studied these scenarios in
the case of muonium to antimuonium conversion, which is
box-diagram dominated and reflects the
underlying process $\mu^+ e^- \to \mu^- e^+$. Here we were
interested in the process $\mu^+ e^- \to e^+ e^-$
which is in general penguin-diagram dominated. We studied this
process both at low and high center-of-momentum energies. At low
(non-relativistic) energy, this corresponds to the muonium decay
$M\to e^+ e^-$. In both scenarios the branching ratios are
exceedingly small (less than $10^{-19}$), basically because,
besides the smallness imposed by a LFV process, there is an extra
factor $\alpha_{em}^3$ arising from the muonium wavefunction. To
avoid this suppression, the alternative is to consider the
collision $\mu^+ e^-$ at higher energies. We find the cross
section for $\mu^+ e^- \to e^+ e^-$ to rise with energy, but
reaching at most $10^{-5}$ fb at $\sqrt s \sim 50$ GeV. If one
expects integrated luminosities not much larger than a few
fb$^{-1}$ at a future muon collider, then this process is not
likely to be observed. Finally, we have pushed to very high
energies, namely above the $W^+W^-$ production threshold. Only in
that case we find that the two scenarios, Model I and Model II,
would induce cross sections for $\mu^+ e^- \to W^+ W^-$ up to the
order of $10^{-1}$ fb or 1 fb. This process, however, has the additional
experimental difficulty that it needs to be distinguished from the
standard background where two neutrinos are present in the final
state.

\acknowledgments C.D. acknowledges support from Fondecyt (Chile)
research grant No. 1030254 and G.C. from Fondecyt grants
No.~1050512 and No.~7010094. The work of C.S.K. was supported in
part by the CHEP-SRC Program and by the Korea Research Foundation
Grant funded by the Korean Government (MOEHRD) No.
KRF-2005-070-C00030. The work of J.D.K was supported by the Korea
Research Foundation Grant (2001-042-D00022).

\begin{appendix}
\section[]{Functions $A_{i j}$ for $e \mu \to WW$ scattering}
\setcounter{equation}{0}

The coefficient functions $A_{i j}(m_i^2,m_j^2,M_W^2,s)$ appearing
in the sum (\ref{sigemuWW1}) are long expressions obtained by
calculating a Trace that follows from squaring the amplitude
(\ref{TmueWW}). For this, we used the program Tracer
\cite{Tracer}. When neglecting $m_{\mu}$ in comparison to $M_W$
and $\sqrt{s}$, the result for $i \not= j$ is
\begin{eqnarray}
A_{i j} &=& \frac{1}{\left[ 48 (m_i - m_j) (m_i + m_j)
M_W^4\right]} {\Big \{} (m_i - m_j) (m_i + m_j) s {\cal K}
\nonumber\\
     && \times
  {\big [} -6(m_i^4 + m_i^2 m_j^2 + m_j^4) + 6 (m_i^2 + m_j^2) M_W^2
         - 24 M_W^4 -  (3 (m_i^2 + m_j^2) - 20 M_W^2) s + s^2 {\big ]}
\nonumber\\
     && - 6 {\big [} (m_i^2 - M_W^2)^2 (m_i^4 + 4 M_W^4) +
           (m_i^3 - 2 m_i M_W^2)^2 s {\big ]} {\cal L}_i
\nonumber\\
 && + 6 {\big [} (m_j^2 - M_W^2)^2 (m_j^4 + 4 M_W^4) +
        (m_j^3 - 2 m_j M_W^2)^2 s {\big ]} {\cal L}_j
{\Big \}} \ ,
\label{Aij}
\\
A_{j j} &=& \frac{1}{48 M_W^4}
 \frac{s {\cal K}}{ \left[ (m_j^2 - M_W^2)^2 + m_j^2 s \right] }
 {\Big [} -24 m_j^8 + 30 m_j^6 (2 M_W^2 - s) +
         m_j^4 (-96 M_W^4 + 68 M_W^2 s - 5 s^2)
\nonumber\\
&&
       + M_W^4 (-48 M_W^4 + 20 M_W^2 s + s^2)
       + m_j^2 (108 M_W^6 - 94 M_W^4 s + 18 M_W^2 s^2 + s^3) {\Big ]}
\nonumber\\
&&
- \frac{1}{8 M_W^4}
 {\Big [} 4 m_j^6 + 4 M_W^4 (-2*M_W^2 + s) +
        m_j^4 (-6 M_W^2 + 3 s) + 2 m_j^2 (5 M_W^4 - 4 M_W^2 s) {\Big ]}
{\cal L}_j
 \ ,
\label{Ajj}
\end{eqnarray}
where we used the short-hand notations
\begin{equation}
{\cal K} = \left[ 1 - \frac{ 4 M_W^2 }{s} \right]^{1/2} \ , \qquad
{\cal L}_k = \ln \left[ \frac{ (2 m_k^2 - 2 M_W^2 + s - s {\cal
K}) }{
         (2 m_k^2 - 2 M_W^2 + s + s {\cal K}) } \right] \ .
\label{calK}
\end{equation}
\end{appendix}

\end{document}